\documentclass{elsart}

\usepackage{graphicx}

\usepackage{amssymb,amsbsy,amsmath}

\newcommand{\op}[1]{%
    \fontdimen12\textfont3=2pt\fontdimen12\scriptfont3=1.4pt%
    \!\null\mathop{\vphantom{#1}\smash{#1}}\limits_{\sim}\null\!}
\newcommand{\xref}[1]{\protect\ref{#1}}
\newcommand{\figref}[1]{Fig.~\protect\ref{#1}}
\newcommand{\fmref}[1]{(\protect\ref{#1})}

\def\ket#1{\, | \, {#1} \, \rangle}

\newcommand{\dint}{\text{d}}

\newcommand {\mofe} {\{$\textrm{Mo}_{72}\textrm{Fe}_{30}$\}}
\newcommand {\mocr} {\{$\textrm{Mo}_{72}\textrm{Cr}_{30}$\}}

\newcommand{\Msat}{{\mathcal M}_{\text{sat}}}

\journal{Polyhedron}

\begin{document}
\begin{frontmatter}

\title{Frustration effects in antiferromagnetic molecules: the
  cuboctahedron} 

\author[adr1]{J\"urgen Schnack\corauthref{cor1}}
\address[adr1]{Universit\"at Bielefeld, Fakult\"at f\"ur Physik,
  D-33501 Bielefeld, Germany}
\corauth[cor1]{Tel: ++49 521 106-6193; fax -6455; Email: jschnack@uni-bielefeld.de}

\author[adr2]{Roman Schnalle}
\address[adr2]{Universit\"at Osnabr\"uck, Fachbereich Physik,
D-49069 Osnabr\"uck, Germany}

\begin{abstract}
Frustration of magnetic systems which is caused by competing
interactions is the driving force of several unusual phenomena
such as plateaus and jumps of the magnetization curve as well as
of unusual energy spectra with for instance many singlet levels
below the first triplet state. The antiferromagnetic
cuboctahedron can serve as a paradigmatic example of certain
frustrated 
antiferromagnets. In addition it has the advantage that its
complete energy spectrum can be obtained up to individual spin
quantum numbers of $s=3/2$ (16,777,216 states).
\end{abstract}


\begin{keyword}
Magnetic Molecules\sep Heisenberg model\sep
Antiferromagnets\sep  Frustration\sep Energy spectrum 

\PACS 
75.50.Ee\sep 75.10.Jm\sep 75.50.Xx\sep 75.40.Mg\sep 24.10.Cn
\end{keyword}
\end{frontmatter}

\section{Introduction}
\label{sec-1}

The magnetism of antiferromagnetically coupled and geometrically
frustrated spin systems is a fascinating subject due to the
richness of phenomena that are observed
\cite{Ram:ARMS94,Gre:JMC01}. Realizations of such systems exist
in one, two, and three dimensions; the most famous being the
two-dimensional kagome lattice
\cite{Gre:JMC01,Diep94,NKH:EPL04,Zhi:PRL02,SHS:PRL02,Atw:NM02}
and the three-dimensional pyrochlore antiferromagnet
\cite{PhysRevLett.88.067203,Moe:CJP01,BrG:Science01,BAA:PRL03,PSS:PRL04,CML:PRL05,Hen:PRL06,SRM:JPA06,ZhT:PRB07}.

\begin{figure}[ht!]
\centering
\includegraphics[clip,width=35mm]{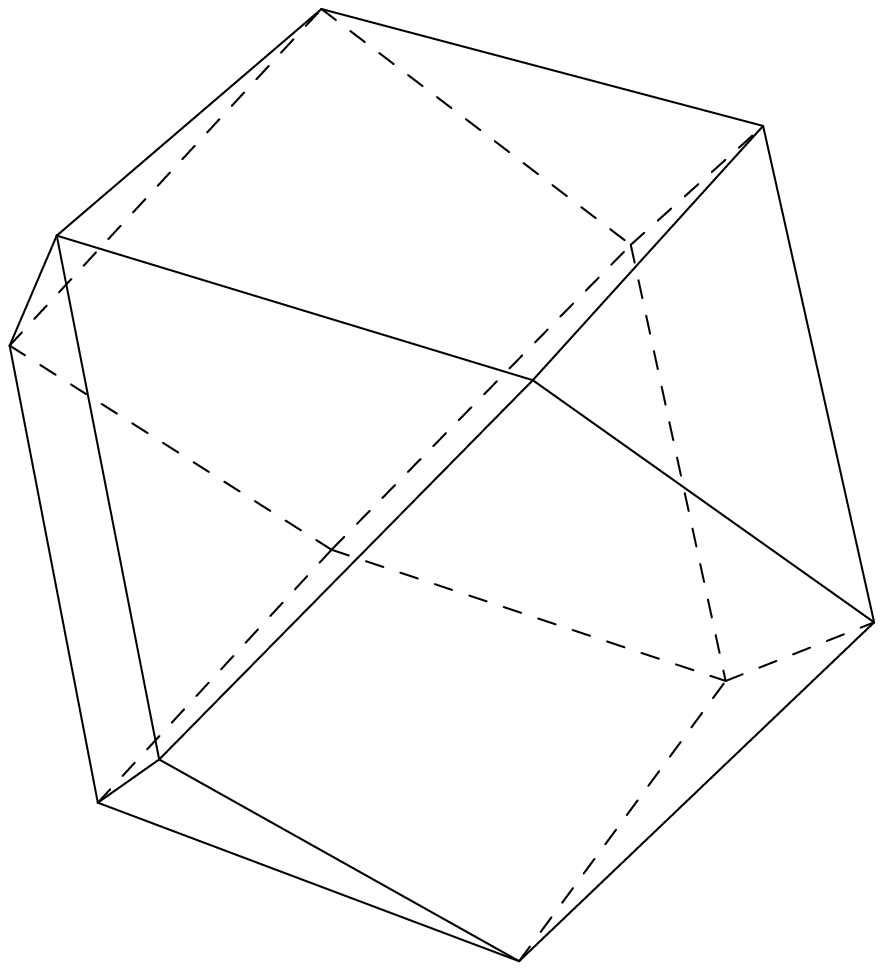}
\quad
\includegraphics[clip,width=35mm]{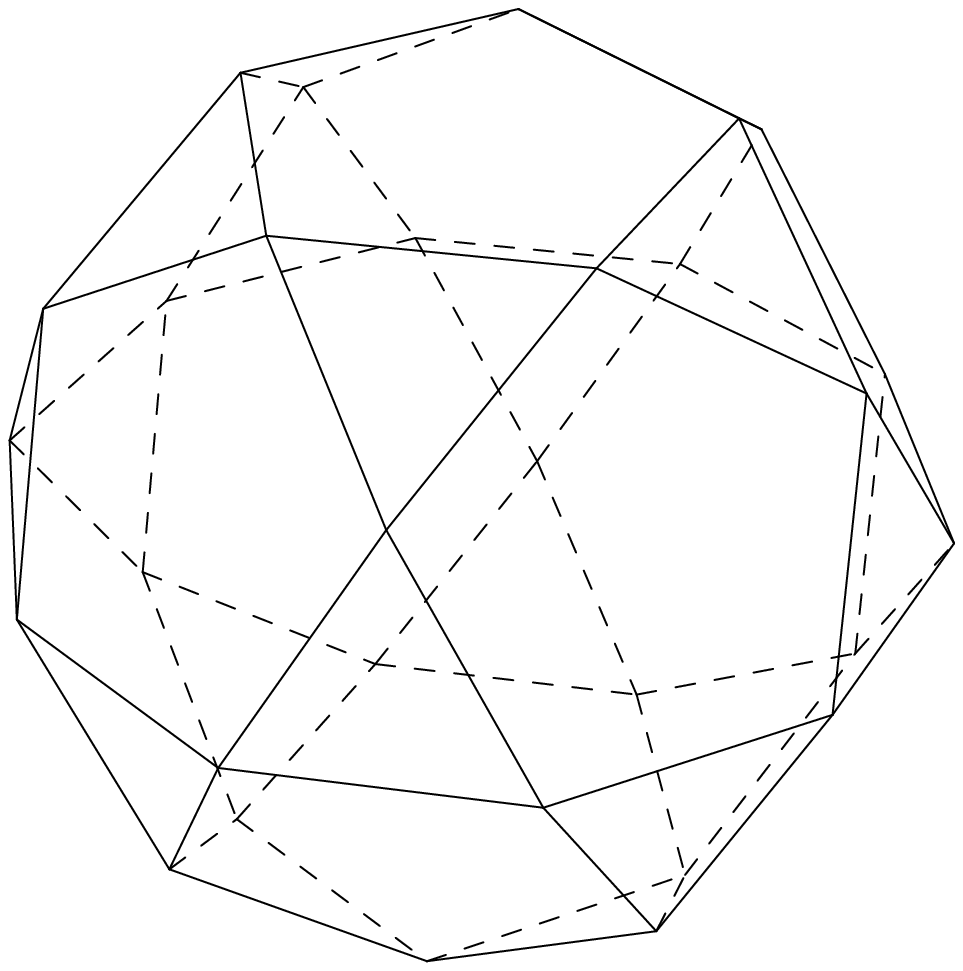}
\quad
\includegraphics[clip,width=45mm]{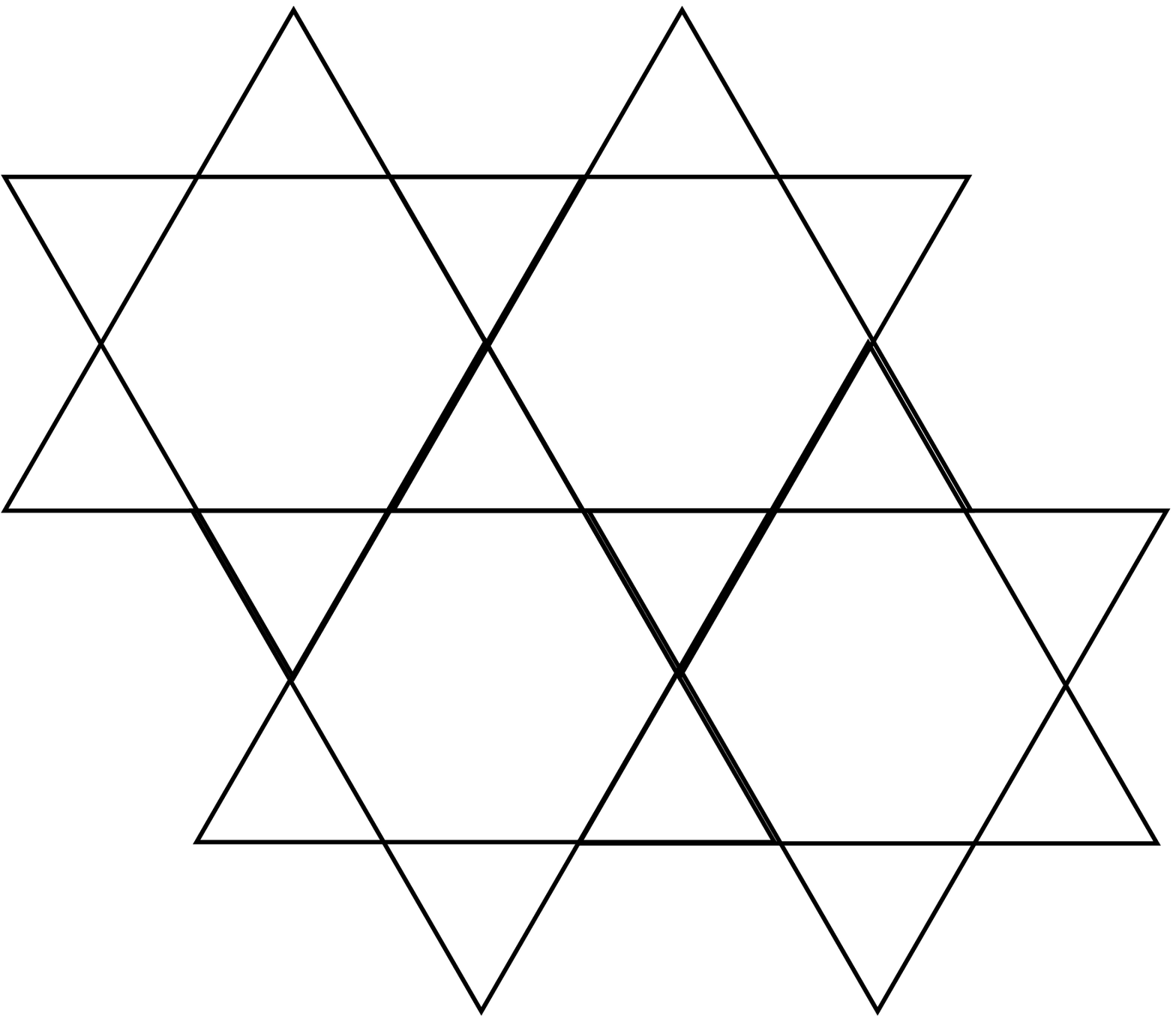}
\caption{Cuboctahedron, icosidodecahedron, and (part of the) kagome lattice}
\label{F-1}
\end{figure}

It is very interesting and from the point of theoretical
modeling appealing that similar but zero-dimensional spin
systems -- in the form of magnetic molecules
\cite{BGG:JCSDT97,MSS:ACIE99,MTS:AC05,TMB:ACIE07,PLK:CC07} --
exist that potentially could show many of the special features
of geometrically frustrated antiferromagnets. Figure~\xref{F-1}
displays the zero-dimensional ``little brothers" of the kagome
antiferromagnet: the cuboctahedron which consists of squares
surrounded by triangles and the icosidodecahedron which consists
of pentagons surrounded by triangles. Such finite size
antiferromagnets offer the possibility to discover and
understand properties that are shared by the infinitely extended
lattices. An example is the discovery of localized independent
magnons \cite{SHS:PRL02,SSR:EPJB01}, which explain the unusual
magnetization jump at the saturation field. Also the plateau at
$1/3$ of the saturation magnetization that appears in systems
built of corner sharing triangles could be more deeply
investigated by looking at the cuboctahedron and the
icosidodecahedron \cite{SNS:PRL05,RLM:PRB08}.

In this article we continue investigations along this line. We
focus on two points. First we discuss the physics of the regular
cuboctahedron as a function of the single spin quantum number
$s=1/2, 1, 3/2$. For these cases all energy eigenvalues could be
obtained with the help of Irreducible Tensor Operator (ITO)
techniques \cite{GaP:GCI93,BCC:IC99,Wal:PRB00} and by
application of point group symmetries. As a second point we
investigate irregular cuboctahedra. This study is motivated by
recent magnetization measurements of the icosidodecahedral
molecules \mofe\ \cite{MSS:ACIE99} and \mocr\ \cite{TMB:ACIE07}
published in Ref.~\cite{SPK:PRB08} which could successfully be
interpreted by a \emph{classical} Heisenberg model with random
antiferromagnetic exchange couplings between the paramagnetic
ions.

\section{Theoretical model}
\label{sec-2}

The physics of many of the mentioned spin systems can be well
understood with the help of the isotropic Heisenberg model,
\begin{eqnarray}
\label{E-2-1}
\op{H}
&=&
- 2\,
\sum_{u<v}\;
J_{uv}\,
\op{\vec{s}}(u) \cdot \op{\vec{s}}(v)
\ .
\end{eqnarray}
Here the sum runs over pairs of spins given by spin operators
$\op{\vec{s}}$ at sites $u$ and $v$. A negative value of the
exchange interaction $J_{uv}$ corresponds to antiferromagnetic
coupling. We refer to a regular body, e.g. cuboctahedron, if
there are only nearest neighbor couplings of constant size
$J$. In the case of an irregular coupling the nearest-neighbor
couplings can assume values according to the chosen
distribution.

Since the Hamiltonian commutes with the total spin, we can find
a common eigenbasis $\{\ket{\nu} \}$ of $\op{H}$, $\op{S}^2$,
and $\op{S}_z$ and denote the related eigenvalues by $E_{\nu}$,
$S_{\nu}$, and $M_{\nu}$, respectively. The eigenvalues of
\fmref{E-2-1} are evaluated in mutually orthogonal subspaces
${\mathcal H}(S,M)$ of total spin $S$ and total magnetic quantum
number $M$ using Irreducible Tensor Operator (ITO) techniques
\cite{GaP:GCI93,BCC:IC99,Wal:PRB00}. In addition point group
symmetries have been applied for the regular cuboctahedron.

\section{The regular cuboctahedron}
\label{sec-3}

The regular cuboctahedron belongs to the class of geometrically
frustrated antiferromagnets built of corner-sharing triangles.
Such systems possess an extended magnetization plateau at $1/3$
of the saturation magnetization $\Msat$ caused by dominant
up-up-down contributions \cite{SNS:PRL05,RLM:PRB08}, an
unusually high jump of the magnetization at the saturation field
due to independent magnons \cite{SSR:EPJB01,SHS:PRL02} as well
as low-lying singlets below the first triplet level
\cite{SSR:JMMM05,SSR:PRB07,RLM:PRB08}. These features are shared
for instance by the icosidodecahedron and by the kagome lattice.

\begin{figure}[ht!]
\centering
\includegraphics[clip,width=45mm]{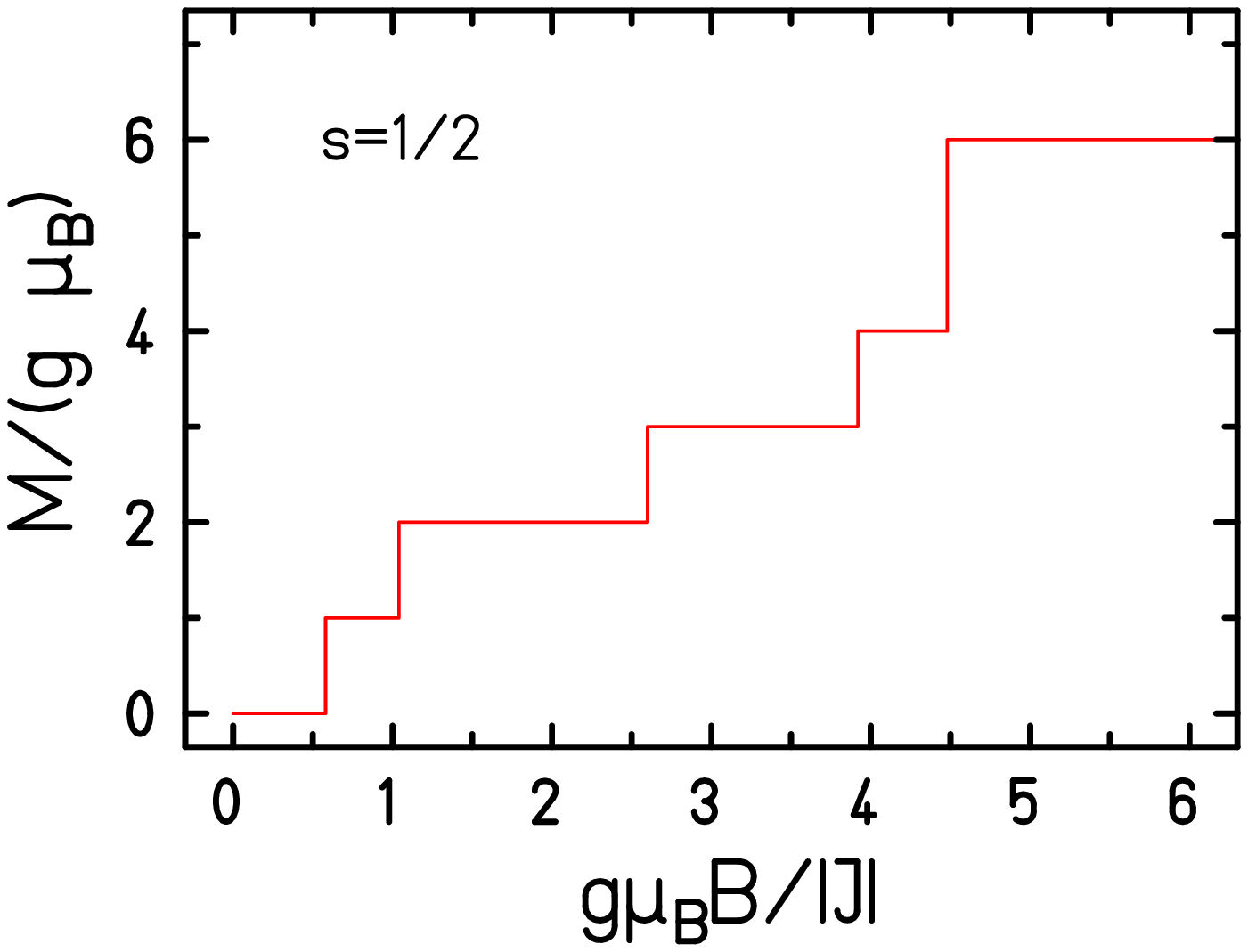}
\includegraphics[clip,width=45mm]{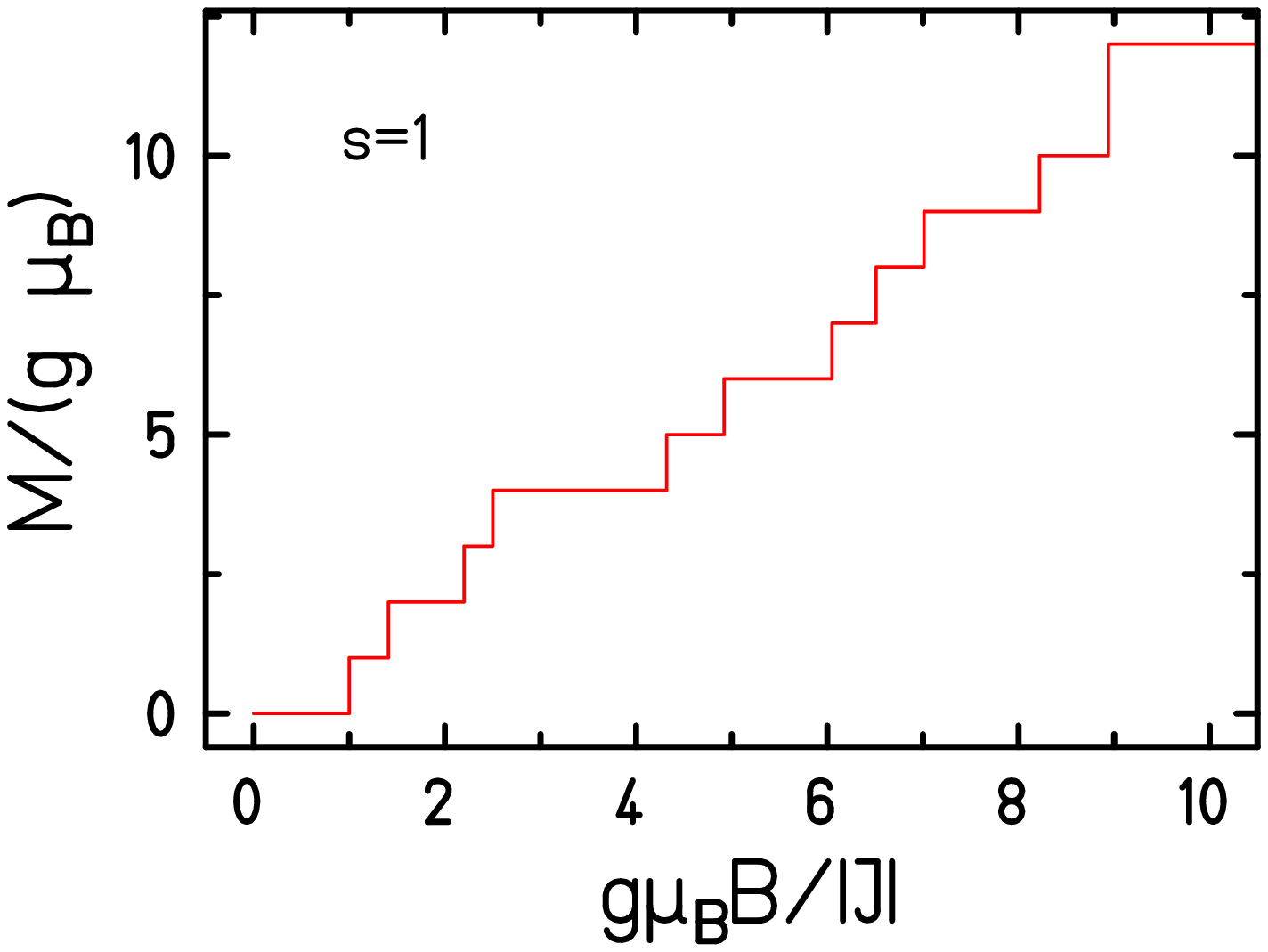}
\includegraphics[clip,width=45mm]{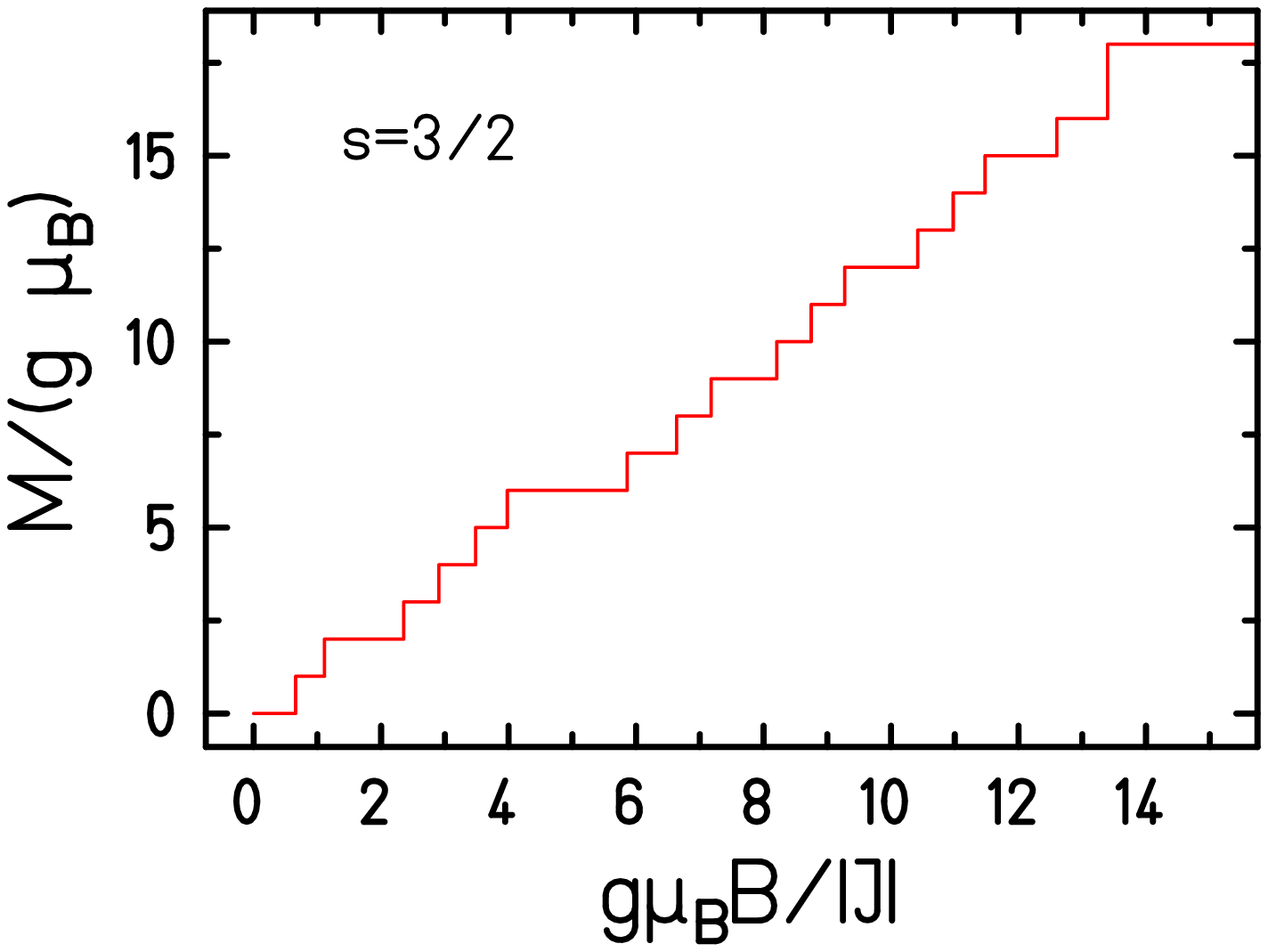}
\caption{Magnetization as a function of applied field at $T=0$
  for the regular cuboctahedron with $s=1/2$, $s=1$, and
  $s=3/2$. The extended plateau at $\Msat/3$ is clearly
  visible.}
\label{F-2}
\end{figure}

\begin{figure}[ht!]
\centering
\includegraphics[clip,width=45mm]{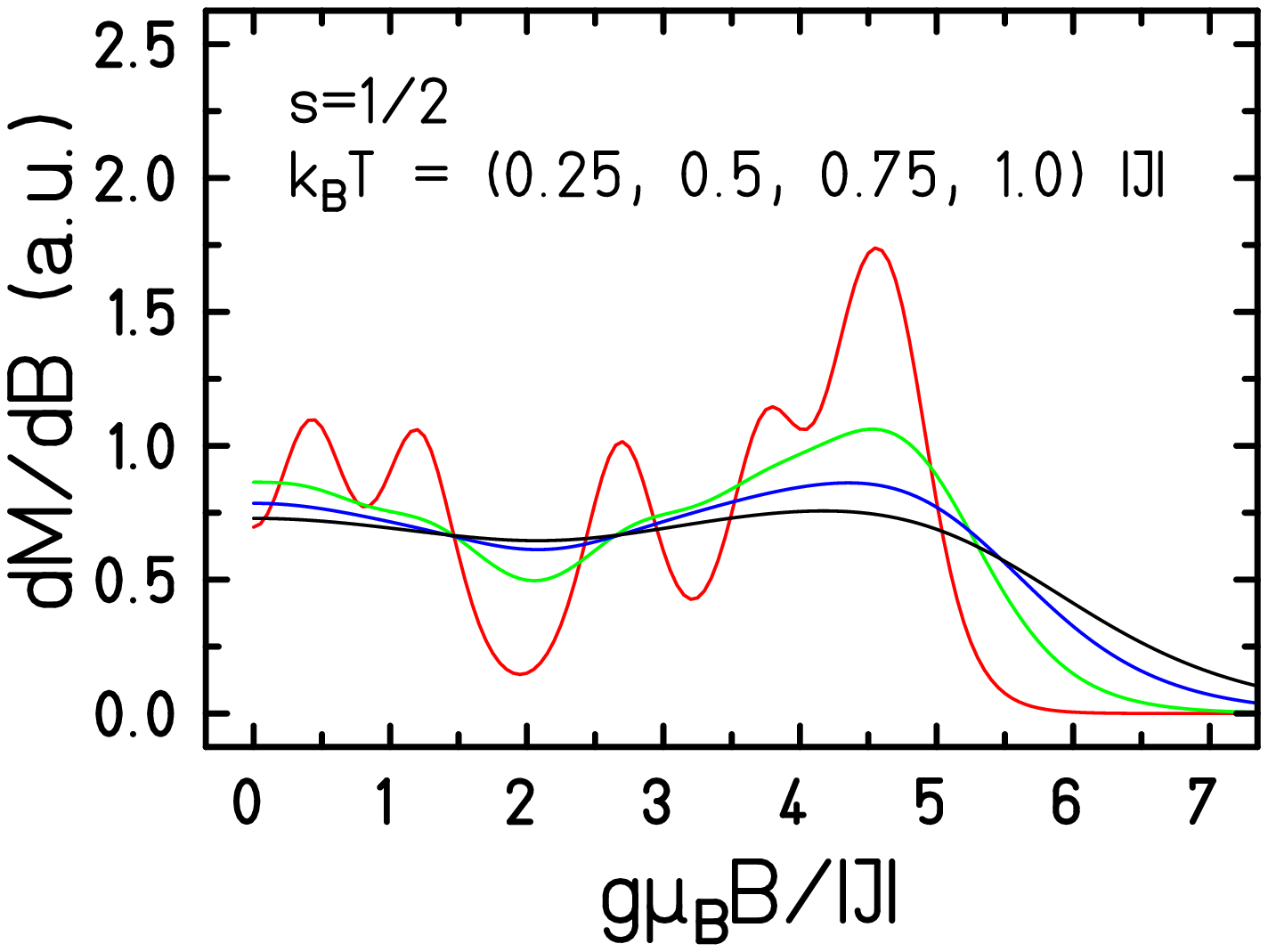}
\includegraphics[clip,width=45mm]{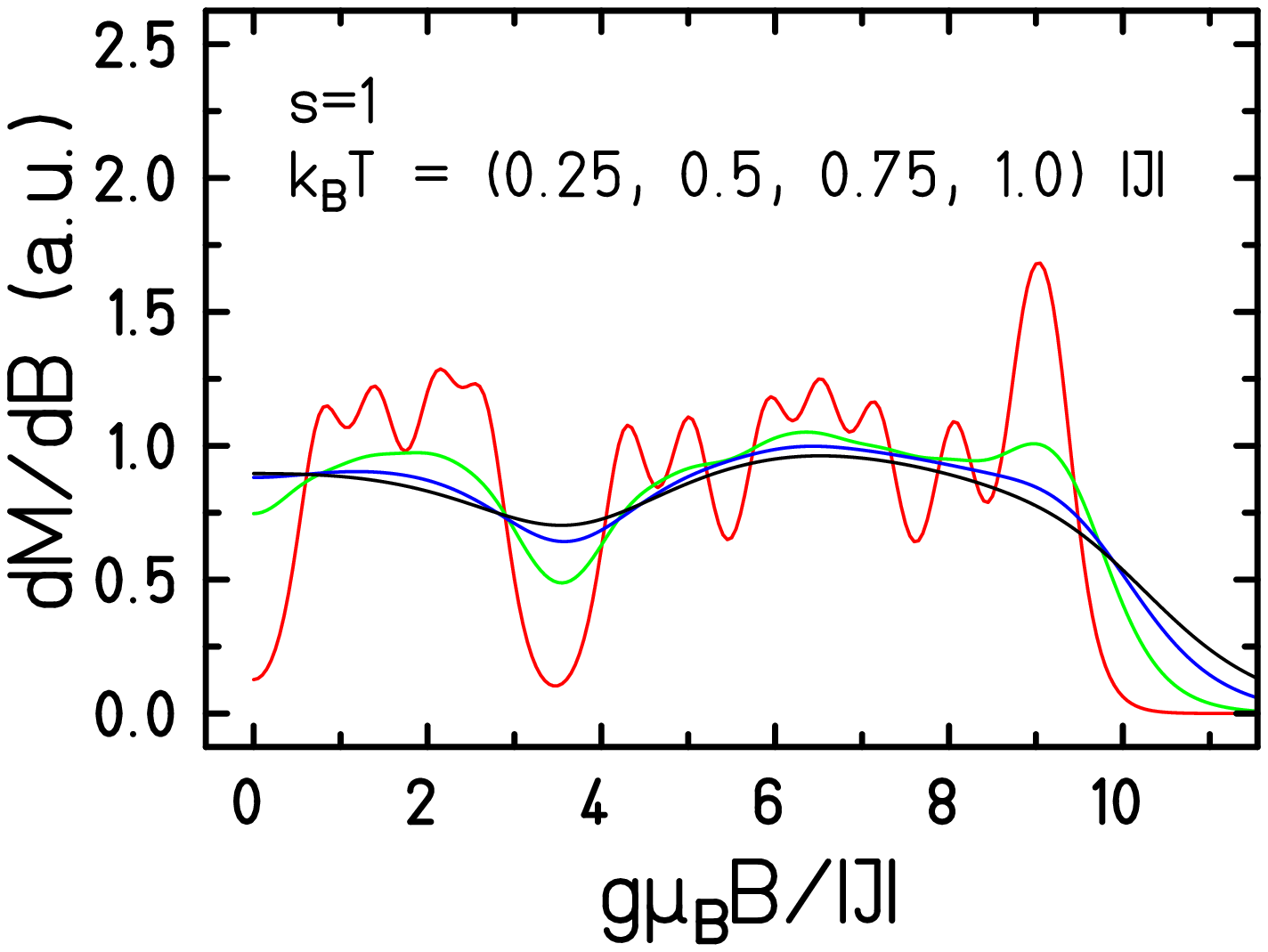}
\includegraphics[clip,width=45mm]{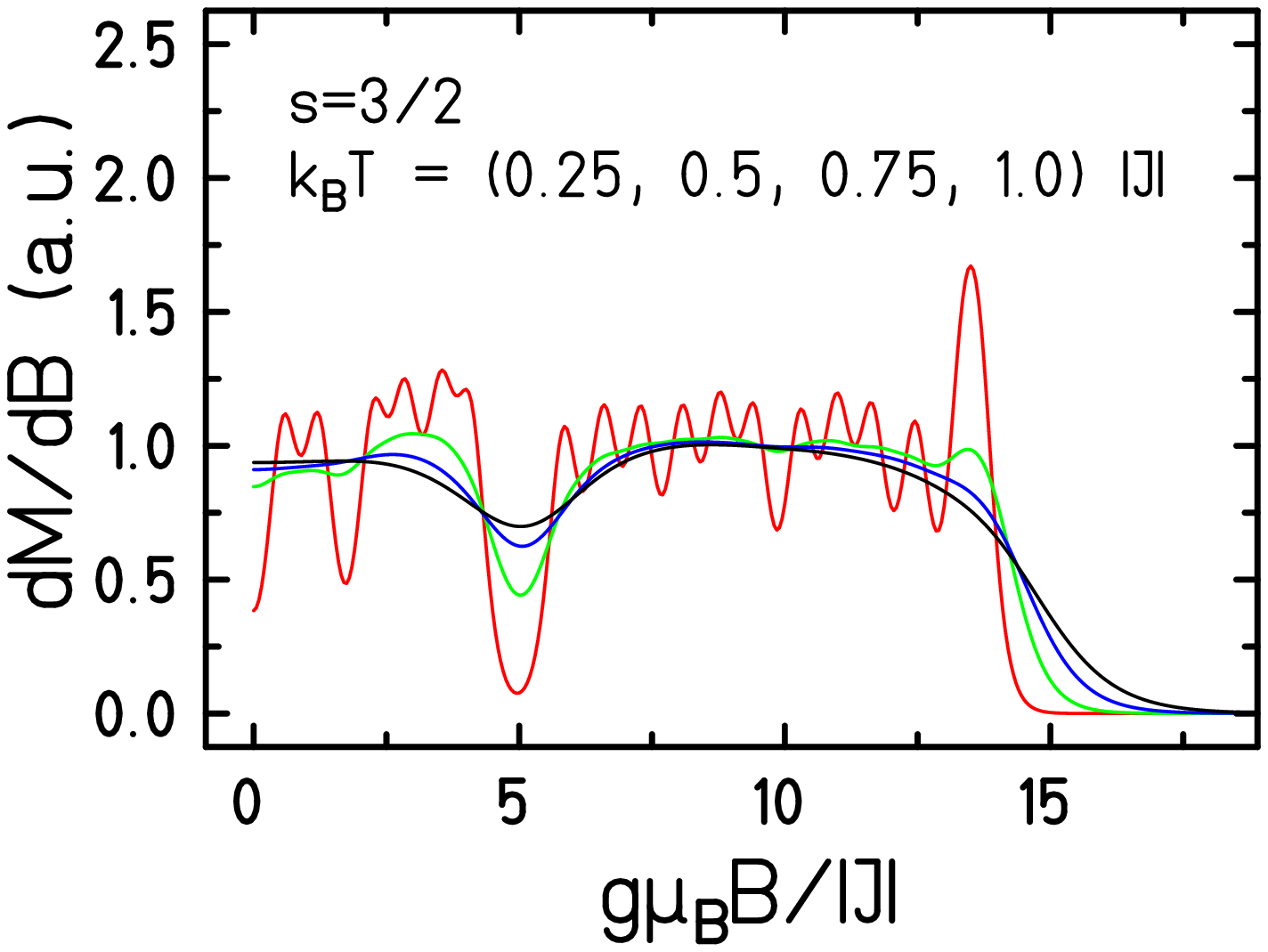}
\caption{Differential susceptibility as a function of applied
  field at $k_BT/|J| = 0.25, 0.5, 0.75, 1.0$ 
  for the regular cuboctahedron with $s=1/2$, $s=1$, and
  $s=3/2$. The smoother the curve, the higher the temperature.}
\label{F-3}
\end{figure}

Figure~\xref{F-2} shows the magnetization curves at $T=0$ for
the regular cuboctahedron with $s=1/2$, $s=1$, and $s=3/2$. These
curves show besides the plateau at $\Msat/3$ a jump to
saturation of height $\Delta M = 2$. Both features are reflected
by the differential susceptibility function which is displayed in
Fig.~\xref{F-3}. Each step in Fig.~\xref{F-2} corresponds to a
peak in Fig.~\xref{F-3}. One notices that the peaks are washed
out for higher temperatures, but that the minimum that
corresponds to the plateau at $\Msat/3$ persists up to
temperatures of the order of the exchange coupling.

As a function of the intrinsic spin $s$ the differential
susceptibility $\dint{\mathcal M}/\dint B$ exhibits two
properties. With increasing spin quantum number $s$ the
individual peaks oscillate more and more with smaller relative
amplitude, but the minimum at $1/3$ is actually sharpened. It
is known that in the classical limit, i.e. for
$s\rightarrow\infty$, the differential susceptibility is
practically flat below the saturation field except for the dip
at $1/3$ \cite{SNS:PRL05}. 

\begin{figure}[ht!]
\centering
\includegraphics[clip,width=45mm]{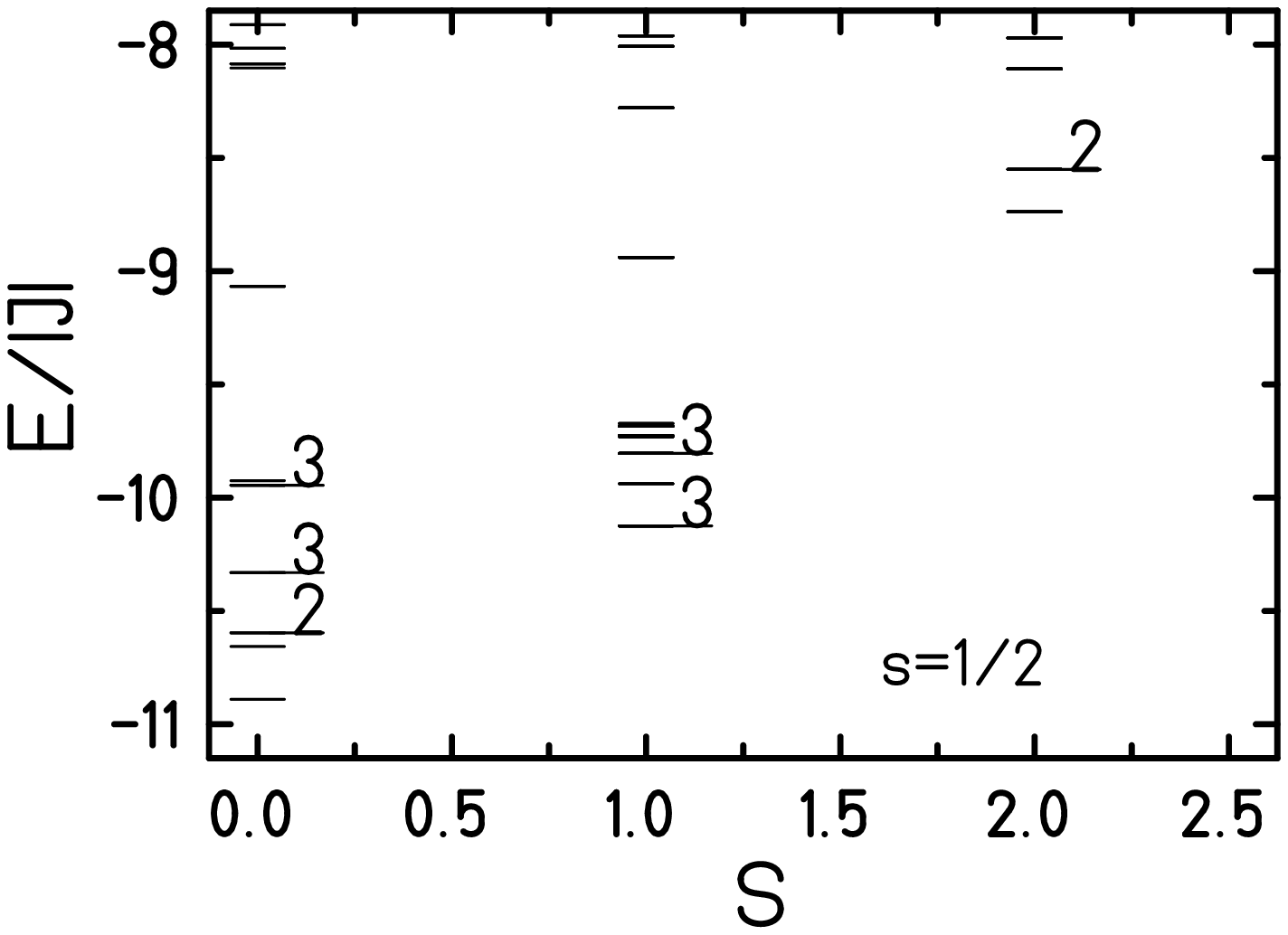}
\includegraphics[clip,width=45mm]{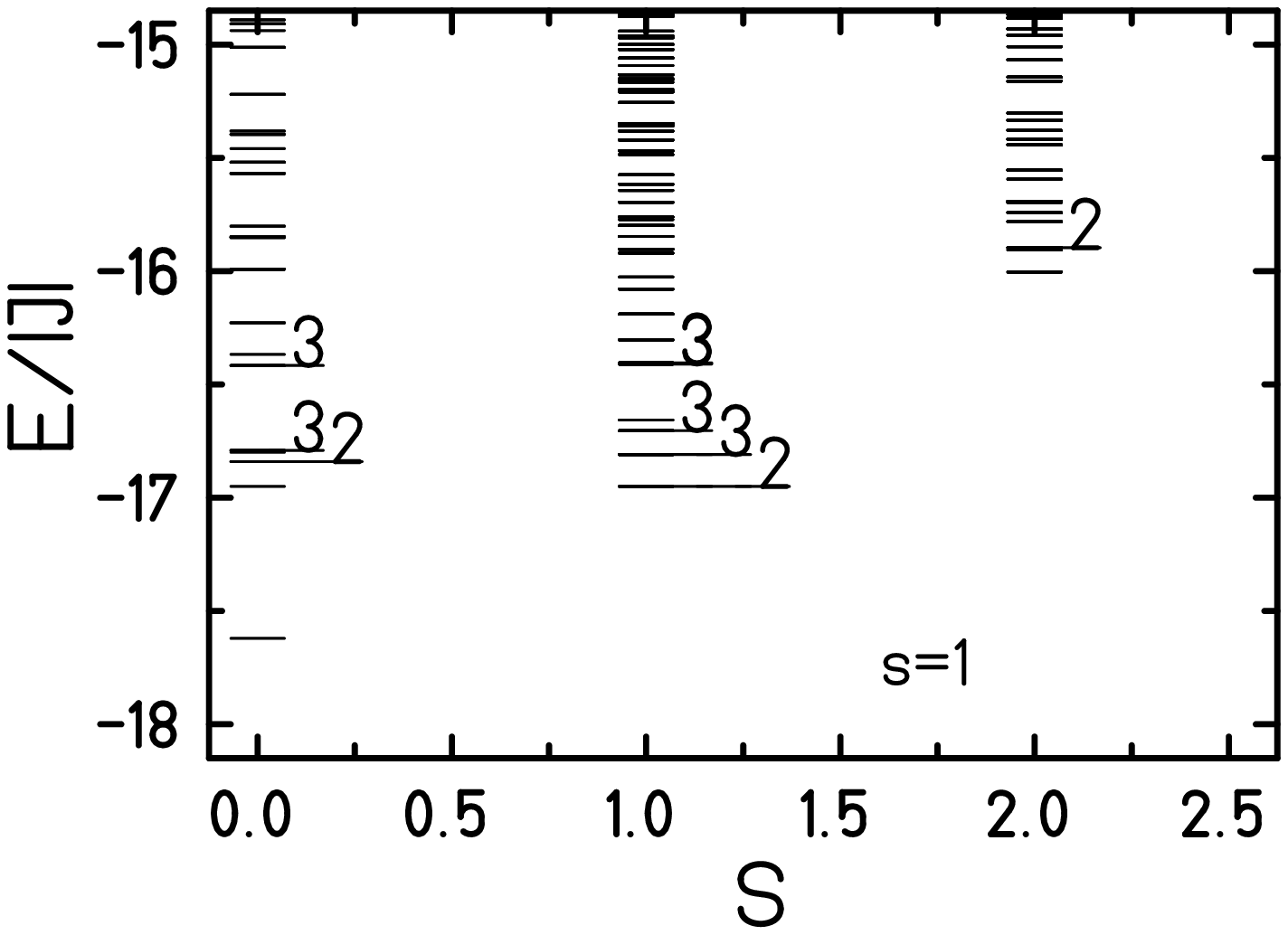}
\includegraphics[clip,width=45mm]{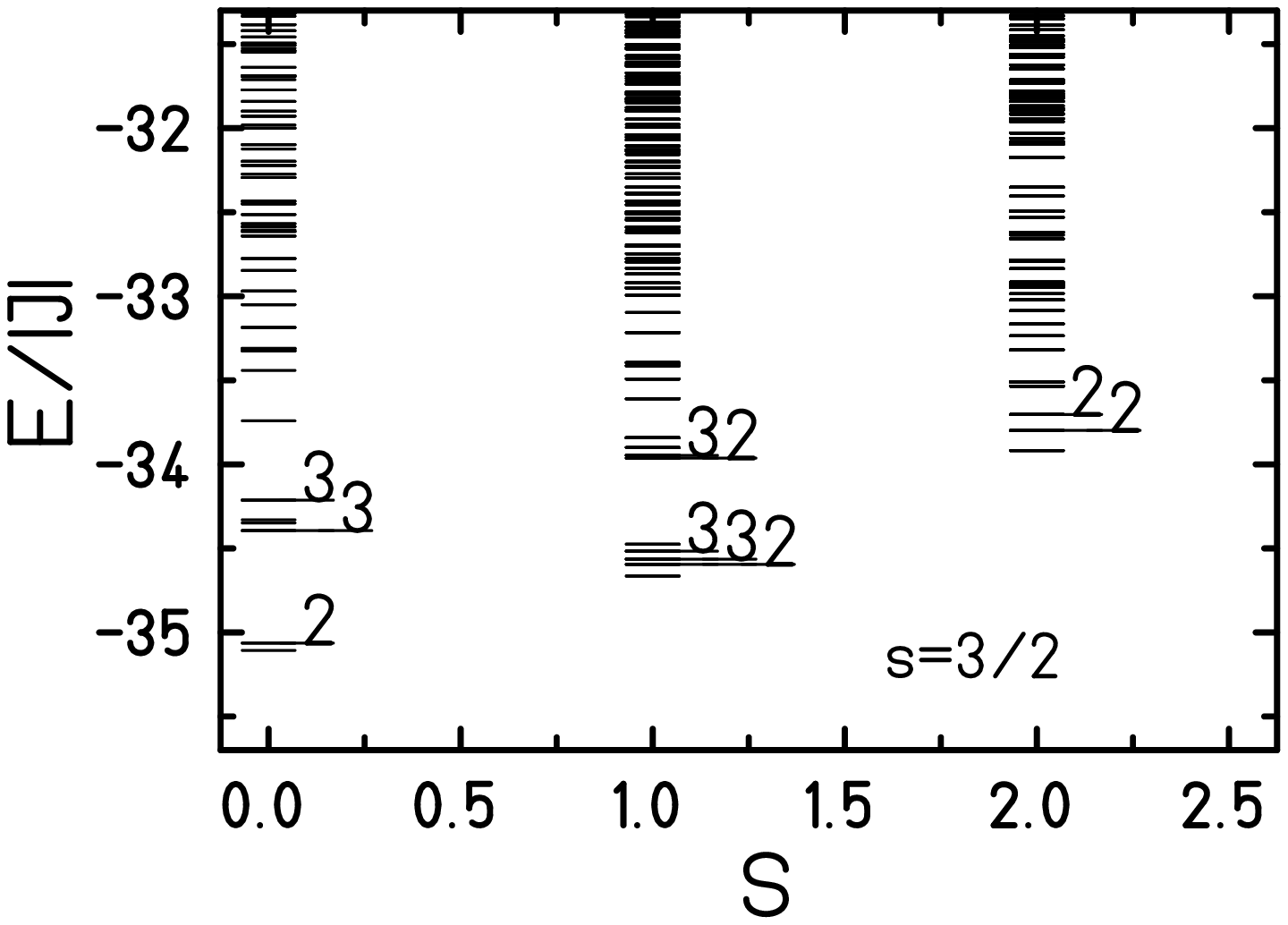}
\caption{Low-lying energy levels for the regular cuboctahedron
  with $s=1/2$, $s=1$, and $s=3/2$. Numbers attached to selected
  levels denote their multiplicities $d_S$; unlabeled levels
  below the highest labeled level have $d_S=1$.}
\label{F-5}
\end{figure}

For zero field \figref{F-5} shows the low-lying energy
levels. In the case of $s=1/2$ (l.h.s. of \figref{F-5}) one
notices the low-lying singlets below the first triplet. These
states are a cornerstone of geometric frustration and as well
present in the kagome lattice and the icosidodecahedron with
$s=1/2$ \cite{SSR:JMMM05}. It is interesting to note that with
increasing $s$, i.e. towards a more classical behavior, the
number of these states decreases. For $s=1$ (middle of
\figref{F-5}) the first excited singlet level is already
(slightly) above the lowest triplet level. For $s=3/2$
(r.h.s. of \figref{F-5}) a doubly degenerate excited singlet
level remains below the lowest triplet, the others have
disappeared. This behavior, i.e. no excited singlets below the
lowest triplet for integer spins and a doubly degenerate excited
singlet below the lowest triplet, does not change anymore for
higher spin quantum numbers as can be checked e.g. by Lanczos
methods.

The rather high symmetry of the cuboctahedron leads to many
degenerate energy levels. As examples we label some low-lying
energy levels in \figref{F-5} by their multiplicity $d_S$,
i.e. by the degeneracy of the whole multiplet. The full
degeneracy including the multiplicity of the magnetic sublevels
$d_M$ is then $d=d_S\times d_M$. Clearly, such high
multiplicities have an important impact on the magnetocaloric
behavior since they increase the entropy for low temperatures
\cite{SSR:PRB07,HoZ:JPCS08}. In the following we would like to
discuss the 
impact of low-lying singlets below the first triplet which in
the case of extended lattices are supposed to condense in
infinite number onto the ground state.

\begin{figure}[ht!]
\centering
\includegraphics[clip,width=45mm]{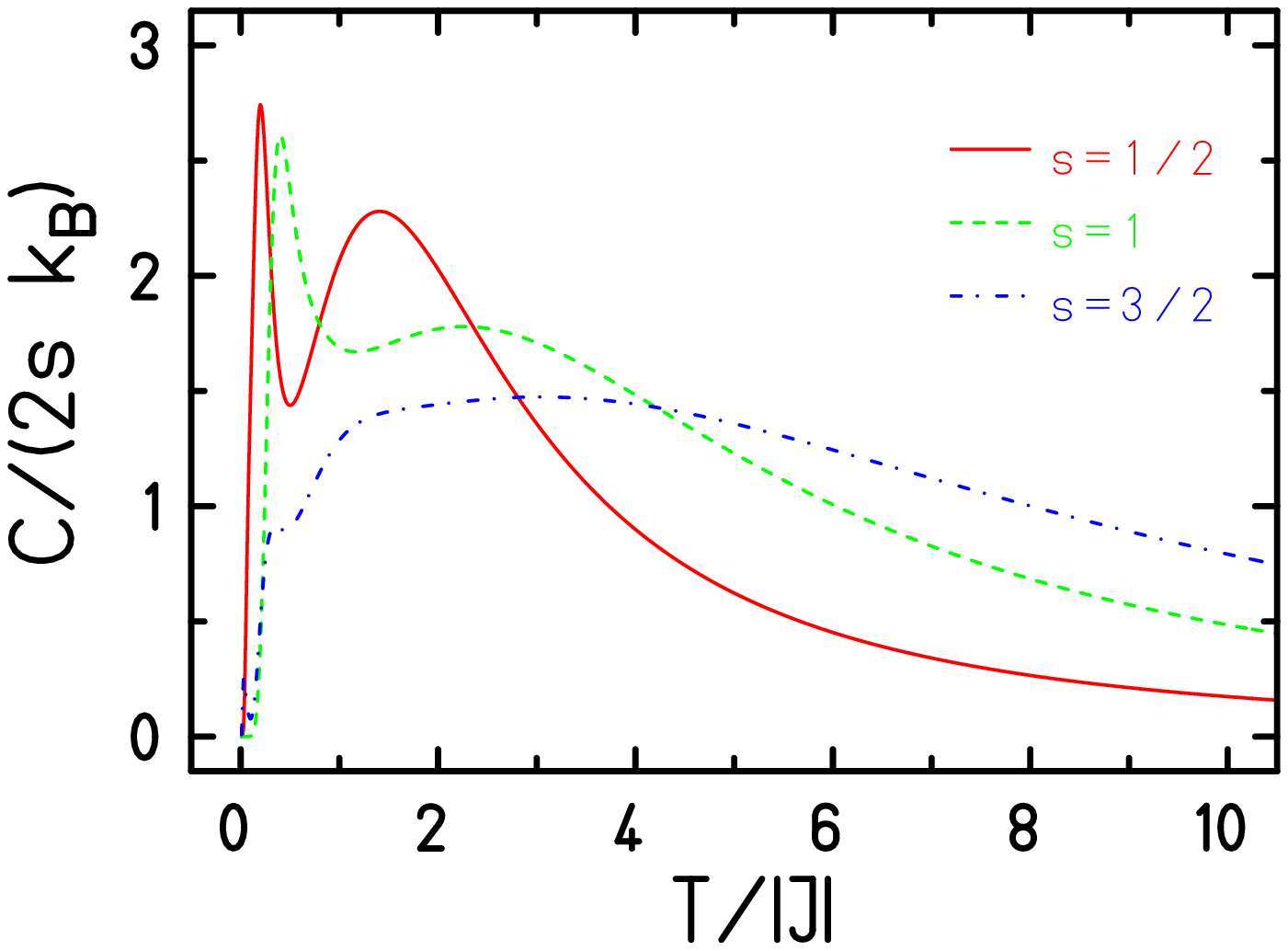}
\quad
\includegraphics[clip,width=45mm]{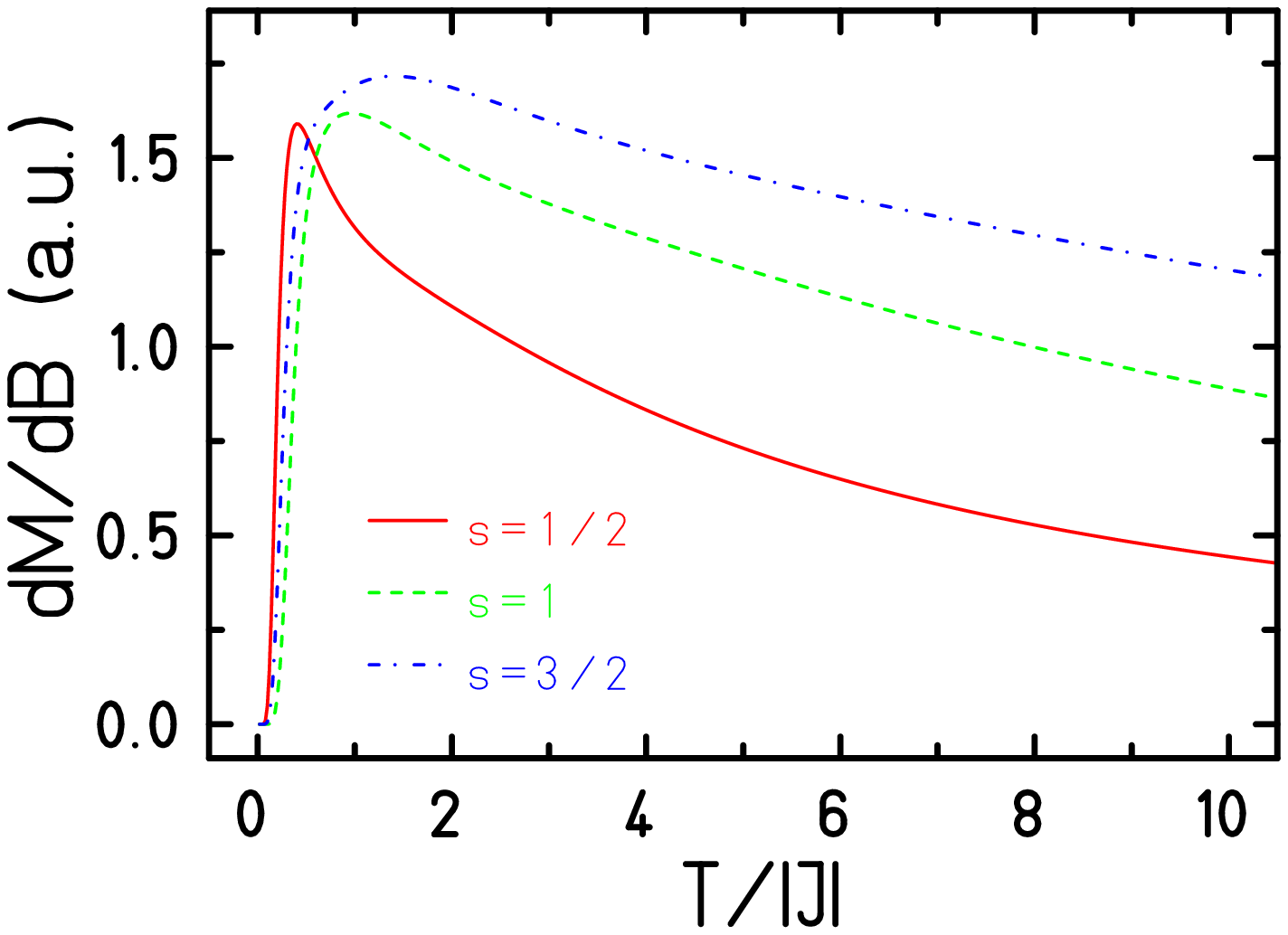}
\caption{Heat capacity (l.h.s.) and zero-field susceptibility
  (r.h.s.) for the regular cuboctahedron with $s=1/2$, $s=1$,
  and $s=3/2$.}
\label{F-4}
\end{figure}

Figure \xref{F-4} compares the heat capacity (l.h.s.) and the
zero-field susceptibility (r.h.s.) for the regular cuboctahedron
with $s=1/2$, $s=1$, and $s=3/2$. The heat capacity shows a
pronounced double peak structure for $s=1/2$ and $s=1$ which
dissolves into a broad peak with increasing spin quantum
number. The broad peak also moves to higher temperatures with
increasing $s$. The reason for the first sharp peak is
twofold. Since there are several gaps between the low-lying
levels the density of states has a very discontinuous structure
which results in the double peak structure.  For $s=1/2$ the
low-lying singlets provide a very low-lying non-magnetic density
of states which is responsible for the fact that the first sharp
peak is at such low temperatures. For $s=1$ the first sharp peak
results from both excited singlet as well as lowest triplet
levels. For $s=3/2$ a remnant of the first sharp peak is still
visible; it is given by the low-lying singlets, but since they
are so few, also influenced by the lowest triplet levels.

The behavior of the heat capacity is contrasted by the
susceptibility on the r.h.s. of \figref{F-4} which reflects
mostly the density of states of magnetic levels and is only
weakly influenced by low-lying singlets. Therefore, the first
sharp peak, or any other structure at very low temperatures, is
absent .

\section{The irregular cuboctahedron}
\label{sec-4}

In this section we investigate how the magnetic properties of
the cuboctahedron change if random variations of the exchange
coupling parameters are introduced. This study is motivated by
recent magnetization measurements of the icosidodecahedral
molecules \mofe\ \cite{MSS:ACIE99} and \mocr\ \cite{TMB:ACIE07}
published in Ref.~\cite{SPK:PRB08}, which were interpreted by
assuming random distributions of exchange parameters in a
classical Heisenberg model description.

\begin{figure}[ht!]
\centering
\includegraphics[clip,width=45mm]{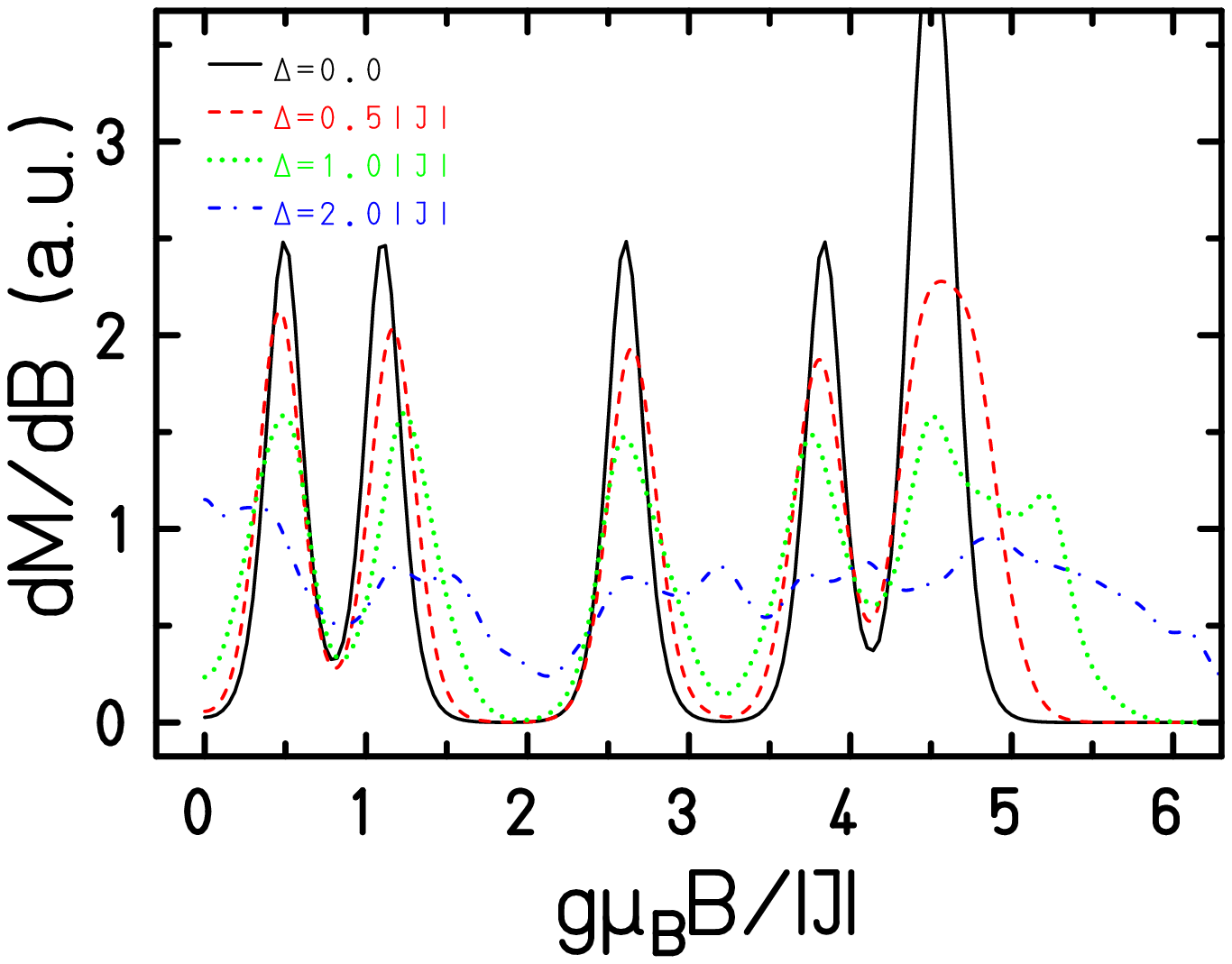}
\includegraphics[clip,width=45mm]{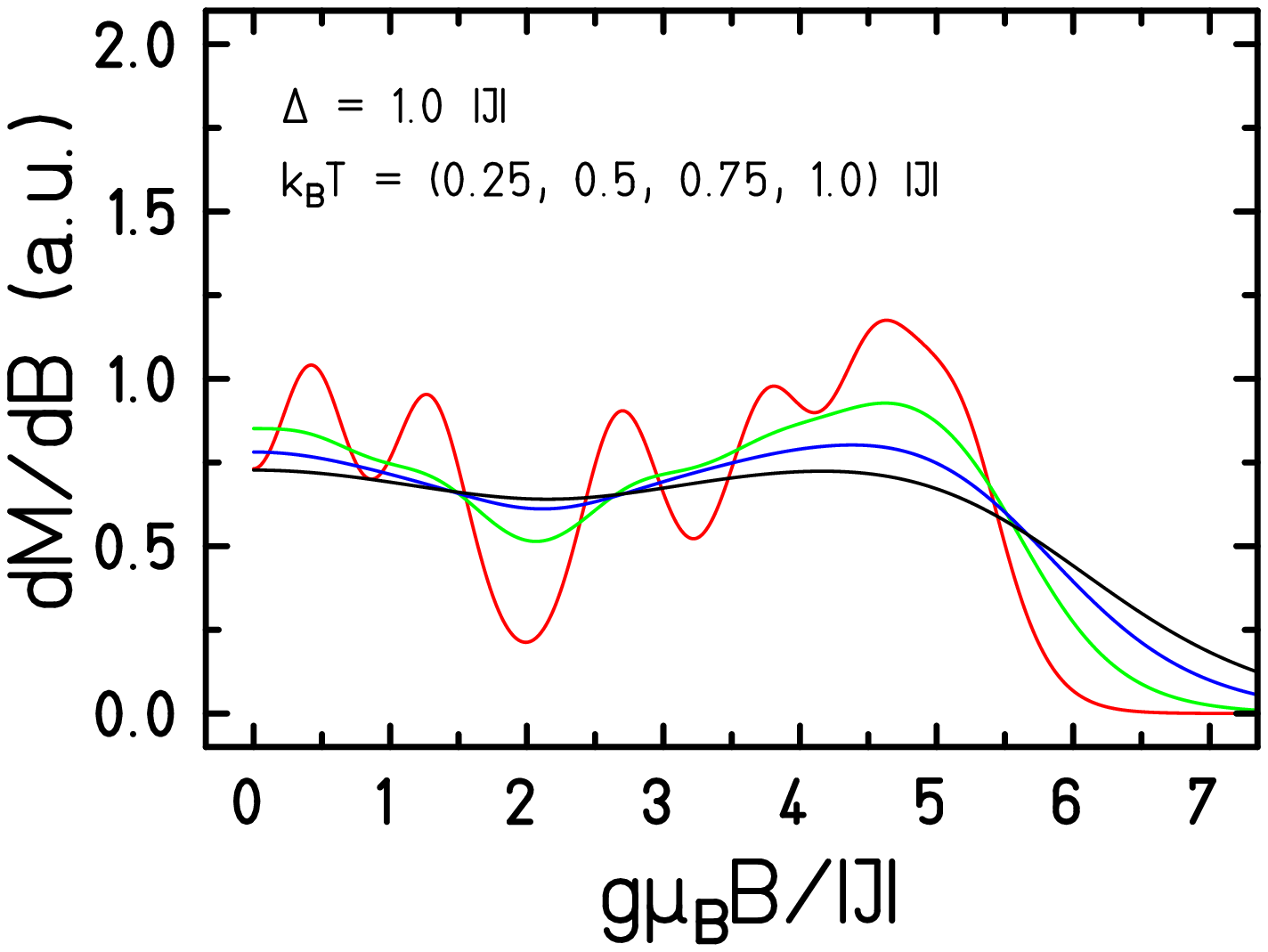}
\includegraphics[clip,width=45mm]{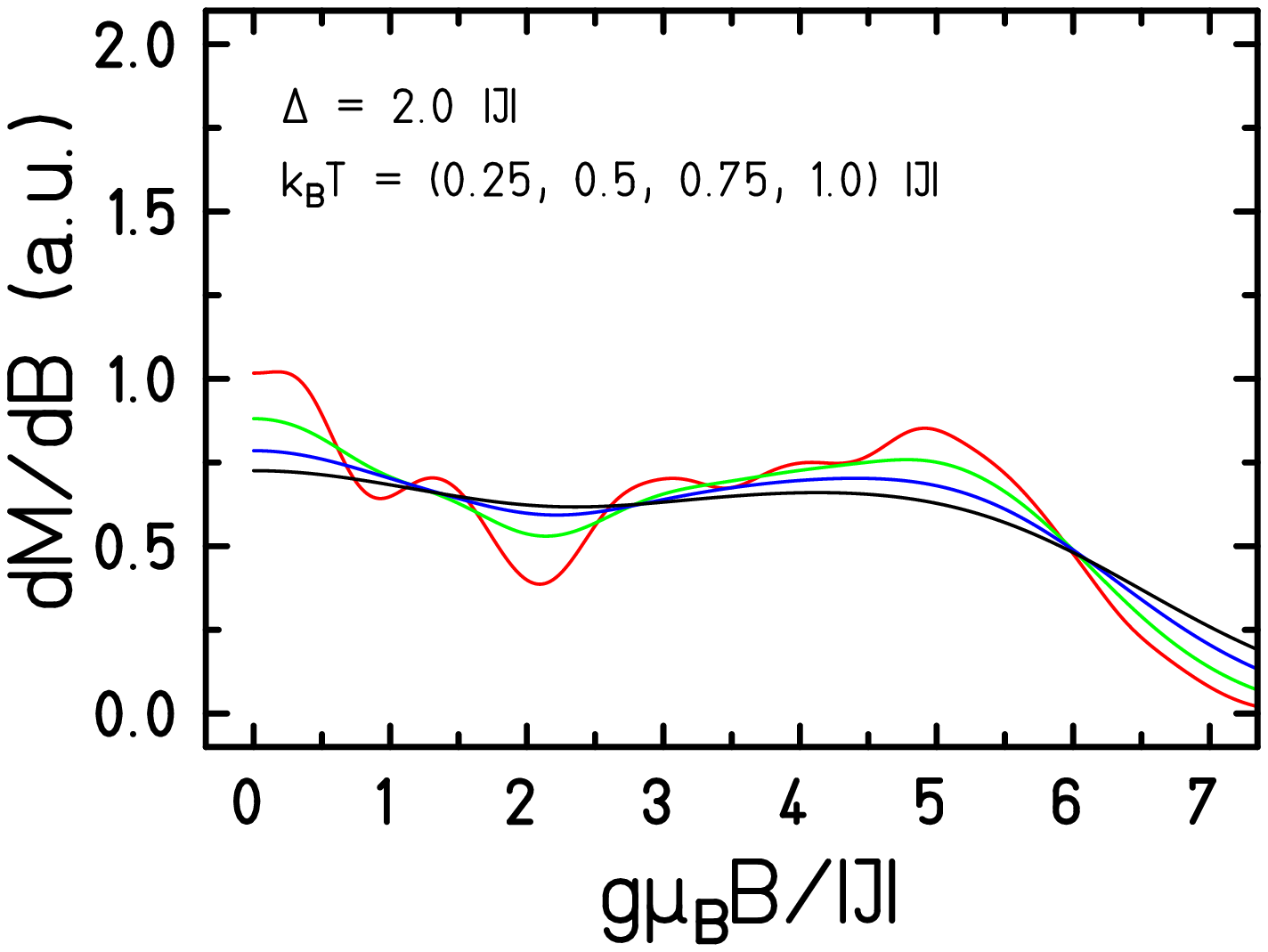}
\caption{Differential susceptibility as a function of applied
  field for the irregular cuboctahedron with $s=1/2$. L.h.s.:
  dependence on the width $\Delta$ of the random distribution.
  Middle: dependence on the temperature $k_BT/|J| = 0.25, 0.5,
  0.75, 1.0$ for $\Delta=1.0 |J|$. R.h.s.: same as middle for
  $\Delta=2.0 |J|$. 
  } 
\label{F-6}
\end{figure}

We introduce variations of the exchange parameters of the
Hamiltonian \fmref{E-2-1} by replacing the common nearest
neighbor exchange parameter $J_{uv}=J$ with values of a flat
random distribution $J-0.5 \Delta \leq J_{uv} \leq J+0.5
\Delta$. Thus the mean exchange parameter is kept to be $J$. In
order to gain sufficient statistical certainty we use ensembles
of 10,000 spectra for realizations of the irregular
cuboctahedron with $s=1/2$; the results do not deviate from
those for 1,000 realizations. For larger $s$ the production of
sufficiently large ensembles is hindered by prohibitively many
diagonalizations of larger matrices.

Figure \xref{F-6} shows the differential susceptibility that
results from averages using distributions with various
$\Delta$. The figure on the l.h.s. compares $\dint{\mathcal
M}/\dint B$ at the rather low temperature of $k_BT=0.1|J|$ for
$\Delta=0$, i.e. the regular cuboctahedron, with $\Delta=0.5
|J|$, $\Delta=1.0 |J|$, and $\Delta=2.0 |J|$. One clearly sees
that the pattern which mainly originates from ground state level
crossings does not change much for $\Delta=0.5 |J|$ and
$\Delta=1.0 |J|$. It needs a variation as large as $\Delta=2.0
|J|$, i.e. ferromagnetic interactions occur, to qualitatively
change the differential susceptibility function. The reason is
that smaller variations do no alter the structure of low-lying
energy gaps. The singlet-triplet gap, which is approximately
$0.765 |J|$, does not vary very much for the ensembles with
smaller $\Delta$, and so does the singlet-triplet crossing which
is determined by the singlet-triplet gap. It needs an
appreciable variance of the exchange parameter distribution in
order to impose large variations of the level crossing fields.

The middle and the r.h.s. of \figref{F-6} display
$\dint{\mathcal M}/\dint B$ for temperatures $k_BT/|J| = 0.25,
0.5, 0.75, 1.0$ and $\Delta=1.0 |J|$ and $\Delta=2.0 |J|$,
respectively. As already explained, there is only very little
difference between the behavior of an irregular cuboctahedron
with $\Delta=1.0 |J|$ (middle) and the regular one. For
$\Delta=2.0 |J|$ (r.h.s.) the differential susceptibility is
much more smeared out which includes an appreciable broadening
at the saturation field.  Considering the irregular
cuboctahedron we can conclude that the magnetic properties are
rather stable against random fluctuations of the exchange
parameters. This means that the striking behavior especially of
the experimental differential susceptibility of \mofe\ and
\mocr\ which shows no signs of level crossings at all
\cite{SPK:PRB08} needs further theoretical exploration of the
microscopic origin.

\section*{Acknowledgment}

Computing time at the Leibniz Computer Center in Garching is
greatly acknowledged as well as helpful openMP advices by Dieter
an Mey and Christian Terboven of High Performance Computing,
RWTH Aachen University. We thank Boris Tsukerblat for motivating
discussions about the Irreducible Tensor Operator technique.



\begin{thebibliography}{10}
\expandafter\ifx\csname url\endcsname\relax
  \def\url#1{\texttt{#1}}\fi
\expandafter\ifx\csname urlprefix\endcsname\relax\def\urlprefix{URL }\fi
\expandafter\ifx\csname selectlanguage\endcsname\relax
  \def\selectlanguage#1{\relax}\fi


\bibitem{Ram:ARMS94}
A.~P. Ramirez, Annu. Rev. Mater. Sci. \textbf{24} (1994) 453

\bibitem{Gre:JMC01}
J.~Greedan, J. Mater. Chem. \textbf{11} (2001) 37

\bibitem{Diep94}
H.~Diep, editor, \emph{Magnetic systems with competing interactions}, World
  Scientific, Singapore (1994)

\bibitem{NKH:EPL04}
Y.~Narumi, K.~Katsumata, Z.~Honda, J.-C. Domenge, P.~Sindzingre, C.~Lhuillier,
  Y.~Shimaoka, T.~C. Kobayashi, K.~Kindo, Europhys. Lett. \textbf{65} (2004)
  705

\bibitem{Zhi:PRL02}
M.~E. Zhitomirsky, Phys. Rev. Lett. \textbf{88} (2002) 057204

\bibitem{SHS:PRL02}
J.~Schulenburg, A.~Honecker, J.~Schnack, J.~Richter, H.-J. Schmidt, Phys. Rev.
  Lett. \textbf{88} (2002) 167207

\bibitem{Atw:NM02}
J.~L. Atwood, Nat. Mater. \textbf{1} (2002) 91

\bibitem{PhysRevLett.88.067203}
O.~Tchernyshyov, R.~Moessner, S.~L. Sondhi, Phys. Rev. Lett. \textbf{88}
  (2002), 6 067203

\bibitem{Moe:CJP01}
R.~Moessner, Can. J. Phys. \textbf{79} (2001) 1283

\bibitem{BrG:Science01}
S.~T. Bramwell, M.~J.~P. Gingras, Science \textbf{294} (2001) 1495

\bibitem{BAA:PRL03}
E.~Berg, E.~Altman, A.~Auerbach, Phys. Rev. Lett. \textbf{90} (2003) 147204

\bibitem{PSS:PRL04}
K.~Penc, N.~Shannon, H.~Shiba, Phys. Rev. Lett. \textbf{93} (2004) 197203

\bibitem{CML:PRL05}
J.-H. Chung, M.~Matsuda, S.-H. Lee, K.~Kakurai, H.~Ueda, T.~J. Sato, H.~Takagi,
  K.-P. Hong, S.~Park, Phys. Rev. Lett. \textbf{95} (2005) 247204

\bibitem{Hen:PRL06}
C.~L. Henley, Phys. Rev. Lett. \textbf{96} (2006) 047201

\bibitem{SRM:JPA06}
H.-J. Schmidt, J.~Richter, R.~Moessner, J. Phys. A: Math. Gen. \textbf{39}
  (2006) 10673

\bibitem{ZhT:PRB07}
M.~E. Zhitomirsky, H.~Tsunetsugu, Phys. Rev. B \textbf{75} (2007) 224416

\bibitem{BGG:JCSDT97}
A.~J. Blake, R.~O. Gould, C.~M. Grant, P.~E.~Y. Milne, S.~Parsons, R.~E.~P.
  Winpenny, J. Chem. Soc.-Dalton Trans.  (1997) 485

\bibitem{MSS:ACIE99}
A.~M\"uller, S.~Sarkar, S.~Q.~N. Shah, H.~B\"ogge, M.~Schmidtmann, S.~Sarkar,
  P.~K\"ogerler, B.~Hauptfleisch, A.~Trautwein, V.~Sch\"unemann, Angew. Chem.,
  Int. Ed. \textbf{38} (1999) 3238

\bibitem{MTS:AC05}
A.~M\"uller, A.~M. Todea, J.~van Slageren, M.~Dressel, H.~B\"ogge,
  M.~Schmidtmann, M.~Luban, L.~Engelhardt, M.~Rusu, Angew. Chem., Int. Ed.
  \textbf{44} (2005) 3857

\bibitem{TMB:ACIE07}
A.~M. Todea, A.~Merca, H.~B{\"o}gge, J.~van Slageren, M.~Dressel,
  L.~Engelhardt, M.~Luban, T.~Glaser, M.~Henry, A.~M{\"u}ller, Angew. Chem.
  Int. Ed. \textbf{46} (2007) 6106

\bibitem{PLK:CC07}
C.~P. Pradeep, D.-L. Long, P.~K{\"o}gerler, L.~Cronin, Chem. Commun.  (2007)
  4254

\bibitem{SSR:EPJB01}
J.~Schnack, H.-J. Schmidt, J.~Richter, J.~Schulenburg, Eur. Phys. J. B
  \textbf{24} (2001) 475

\bibitem{SNS:PRL05}
C.~Schr{\"o}der, H.~Nojiri, J.~Schnack, P.~Hage, M.~Luban, P.~K{\"o}gerler,
  Phys. Rev. Lett. \textbf{94} (2005) 017205

\bibitem{RLM:PRB08}
I.~Rousochatzakis, A.~M. L\"{a}uchli, F.~Mila, Phys. Rev. B \textbf{77} (2008)
  094420

\bibitem{GaP:GCI93}
D.~Gatteschi, L.~Pardi, Gazz. Chim. Ital. \textbf{123} (1993) 231

\bibitem{BCC:IC99}
J.~J. Borras-Almenar, J.~M. Clemente-Juan, E.~Coronado, B.~S. Tsukerblat,
  Inorg. Chem. \textbf{38} (1999) 6081

\bibitem{Wal:PRB00}
O.~Waldmann, Phys. Rev. B \textbf{61} (2000) 6138

\bibitem{SPK:PRB08}
C.~Schr\"{o}der, R.~Prozorov, P.~K\"{o}gerler, M.~D. Vannette, X.~Fang,
  M.~Luban, A.~Matsuo, K.~Kindo, A.~M\"{u}ller, A.~M. Todea, Phys. Rev. B
  \textbf{77} (2008) 224409

\bibitem{SSR:JMMM05}
R.~Schmidt, J.~Schnack, J.~Richter, J. Magn. Magn. Mater. \textbf{295} (2005)
  164

\bibitem{SSR:PRB07}
J.~Schnack, R.~Schmidt, J.~Richter, Phys. Rev. B \textbf{76} (2007) 054413

\bibitem{HoZ:JPCS08}
A. Honecker, M. E. Zhitomirsky, J. Phys. Conf. Ser., in print;
arXiv:0809.4414v1

\end{thebibliography}

\end{document}